\documentclass[12pt]{iopart}
\pdfoutput=1
\expandafter\let\csname equation*\endcsname\relax
\expandafter\let\csname endequation*\endcsname\relax
\usepackage[numbers,sort&compress]{natbib}
\usepackage{amsmath,amssymb,eucal,graphicx,subfigure}
\usepackage{url}
\usepackage{bm}
\usepackage{color}

\begin{document}

\title{Simple Parking Strategies}
\author{P. L. Krapivsky}
\address{Department of Physics, Boston University, Boston, MA, 02215 USA}
\author{S. Redner}
\address{Santa Fe Institute, 1399 Hyde Park Road, Santa Fe, NM, 87501 USA}

\begin{abstract}
  We investigate simple strategies that embody the decisions that one faces
  when trying to park near a popular destination.  Should one park far from
  the target (destination), where finding a spot is easy, but then be faced
  with a long walk, or should one attempt to look for a desirable spot close
  to the target, where spots may be hard to find?  We study an idealized
  parking process on a one-dimensional geometry where the desired target is
  located at $x=0$, cars enter the system from the right at a rate $\lambda$
  and each car leaves at a unit rate.  We analyze three parking
  strategies---meek, prudent, and optimistic---and determine which is
  optimal.

\end{abstract}


\section{Introduction}

When driving to a popular destination, nearby parking spots are hard to find.
Where should one park?  Should one park far from the destination, or target,
where spaces are likely to be plentiful and then walk a long way to the
target?  Alternatively, should one be optimistic and drive close to the
target and look only for nearby parking?  If one uses the latter strategy, it
is possible that there are no nearby parking spots and then one has to
backtrack to find a more distant parking spot, thereby wasting time.

\begin{figure}[ht]
  \centerline{\includegraphics[width=0.65\textwidth]{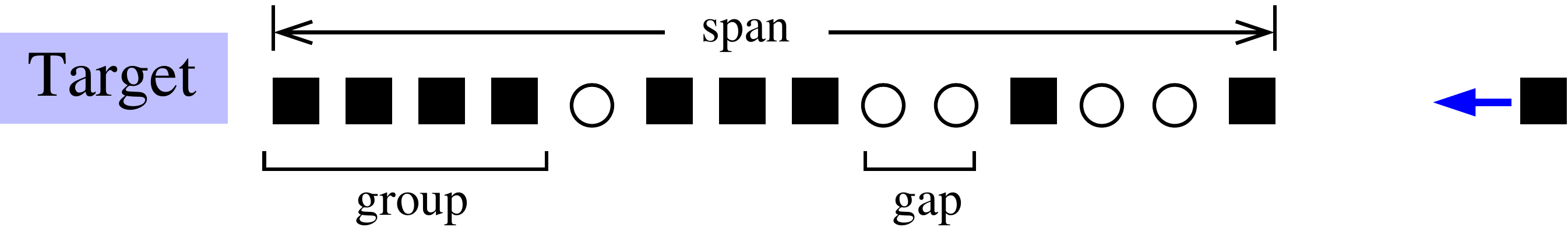}}
  \caption{Parking in a one-dimensional lot where cars (squares) enter from
    the right.  Circles represent empty spots, and empty spots to the right
    of the furthest car are not shown.  The spatial range of the parked cars
    is defined as the span.}
\label{cartoon}  
\end{figure}

As one might anticipate, this practical problem has been the focus of
considerable study in the transportation engineering literature (see, e.g.,
\cite{V82,YTT91,AP91,TR98,AR99,TL06,KLW14} and references therein).  These
practically minded studies include many real-world effects, such as parking
costs, parking limits, and urban planning implications, that cannot be
accounted for in minimalist physics-based modeling.  In the context of
granular compaction, the ``parking lot model'' describes how a finite
interval with input and output of cars progressively densifies due to various
compaction mechanisms~\cite{R58,E93,KB94,KNT99,TTV01,WH03,KRB10}.  In this
work, we explore simple parking strategies in an idealized one-dimensional
geometry and determine their relative advantages.

In our modeling, we assume that cars enter the system from the right at a
fixed rate $\lambda$ and park at integer points along a one-dimensional
semi-infinite line, which plays the role of a parking lot.  Cars also
independently leave the lot at a unit rate.  For a lot that contains $N$
parked cars, the total car departure rate therefore equals $N$.  The first
car that enters will park at $x=1$, the closest spot to the target.  The
second car will park at $x=2$ if the first car has not left at the moment
when the second car arrives.  We assume that all cars have the same size and
fill exactly one parking spot.  As the parking lot fills, the parked cars
forms contiguous groups that are interspersed with gaps (Fig.~\ref{cartoon}).
The most interesting situation is $\lambda\gg 1$, so that the number of
parked cars is large.  We also assume that when a
car enters the lot, it has time to find a parking space before the next car
enters.

In the steady state, the number of parked cars is a random quantity that
fluctuates around its average value that is equal to $\lambda$.  In this
state, cars enter and leave the lot at the same rate, but the spatial
distribution of parked cars is continually rearranging.  This situation has
some commonality of continually writing and erasing files on a computer disk.
As the disk becomes more full, one faces the problem of disk fragmentation,
which can significantly degrade its performance (see, e.g., \cite{K72}).  In
our parking model, all cars occupy a single open spot and the problem of
parking lot ``fragmentation'' is generally minor.

At first sight, our parking problem resembles optimal stopping problems for
which a vast literature exists (see, e.g.,
\cite{Miller,D70,Chow71,BG84,F89,OS}).  A crucial aspect of our work,
however, is that the spatial distribution of parked cars depends on the
strategy, while in optimal stopping problems these occupancies are random.
Another distinction is that, according to our rules, the driver cannot 'see'
the state of closer parking spots (except for a contiguous string of open
spots if the current spot is open).

In the next section, we outline the three parking strategies that will be
studied in this work.  We then turn to the dynamics of the number of parked
cars, which does not depend on the parking strategy.  In Sec.~\ref{sec:meek},
we determine the spatial distribution of cars in the strategy where all
drivers are meek and park behind the first car encountered. There is a
mapping between this strategy and a model of microtubule dynamics~
\cite{AKRMC07}, so we describe the mapping and present some asymptotic
results from \cite{AKRMC07}.  We next turn to two more realistic strategies
that we term as ``optimistic'' and ``prudent'' and determine their relative
merits.

\section{Parking Strategies}
\label{sec:strategies}

\begin{figure}[ht]
  \centerline{\subfigure[]{\includegraphics[width=0.55\textwidth]{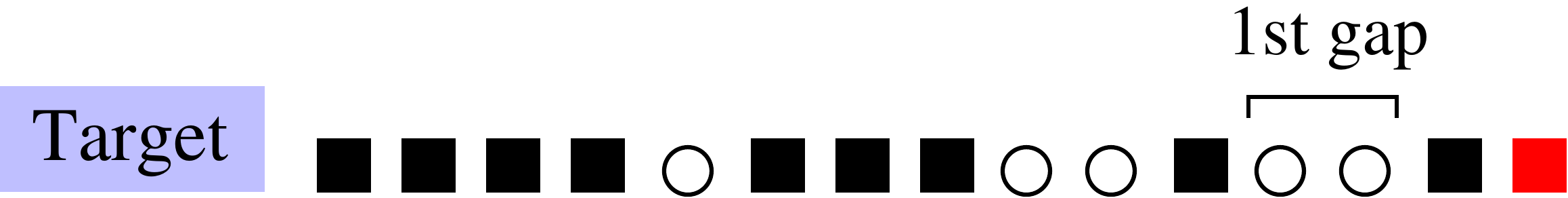}}}
  \centerline{\subfigure[]{\includegraphics[width=0.55\textwidth]{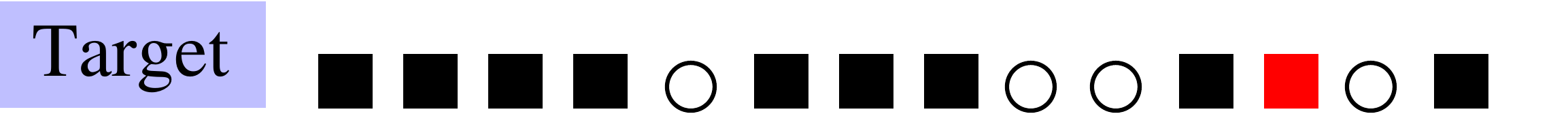}}}
  \centerline{\subfigure[]{\includegraphics[width=0.55\textwidth]{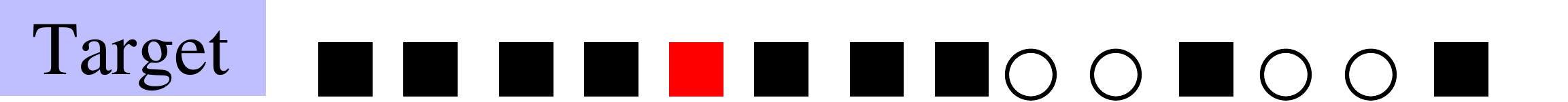}}}
  \caption{Illustration of different parking strategies for the same state of
    the parking lot: (a) meek, (b) prudent, and (c) optimistic.  The red
    square denotes the newly parked car. }
\label{add}  
\end{figure}

Cars arrive one at a time at rate $\lambda$, and each arriving car parks at
one of the available parking spots.  We postulate that the drivers have no
information about available spots; otherwise they would go straight to the
closest available spot.  With this uncertainty, there are various natural
parking strategies.  We analyze three  strategies (Fig.~\ref{add}):
\begin{enumerate}
\item \textbf{Meek:} Park at the first available spot just behind the
  rightmost parked car.

\item\textbf{Prudent:} Go to the first gap and park at the left end of this
  gap.  If there are no vacancies, go all the way to $x=0$, then backtrack,
  and finally park behind the rightmost parked car.

\item {\bf Optimistic:} Go all the way to $x=0$ and then backtrack to the
  closest available spot.  If there are no vacancies, this backtracking ends
  by parking behind the rightmost parked car.
\end{enumerate}

The meek driver wastes no time looking for a parking spot and just parks at
the first available spot that is behind the most distant parked car.  This
strategy is risibly inefficient; many good parking spots are unfilled and
most cars are parked far from the target.  The prudent driver bets that there
is at least one vacancy in the lot.  If this bet is wrong, the prudent  driver
wastes the time to travel to $x=0$ and then backtracks to where it would have
parked by employing the meek strategy.  The optimistic driver bets that there
is a spot close to the target and thus drives to the target and parks at the
first vacancy encountered by backtracking (Fig.~\ref{add}(c)).  If a vacancy
does not exist, the optimistic driver must also backtrack and park at the end
of the line of parked cars.

\section{Dynamics of the Number of Cars}

A basic characterization of this parking process is the total number $N(t)$
of parked cars at time $t$.  If we ignore the time spent in actually parking,
the random variable $N(t)$ is independent of the parking strategy.  The
probability distribution $P_N(t)$ that there are $N$ parked cars at time $t$
satisfies the master equation
\begin{align}
\label{PNdot}
\frac{dP_N}{dt}= \lambda P_{N-1} + (N+1)P_{N+1}  - (\lambda+N)P_N\,.
\end{align}
The first term on the right accounts for the gain in $P_N$ because a car
parks in a lot with $N-1$ cars, the second term accounts for the gain in
$P_N$ because a car leaves when the lot contains $N+1$ cars, and the last
term accounts for the loss of $P_N$ because either a new car parks or a car
leaves when the lot contains $N$ cars.

The solution to Eq.~\eqref{PNdot} can be obtained by the generating function
method for an arbitrary initial condition (the derivation is given, e.g., in
Ref.~\cite{AKRMC07}).  If the parking lot is initially empty,
$P_N(0)=\delta_{N,0}$, the distribution of the number of parked cars is given
by the Poisson distribution
\begin{equation}
\label{Poisson}
P_N(t) = \frac{\big[\lambda(1-e^{-t})\big]^N}{N!}\, e^{-\lambda(1-e^{-t})}\,\,
\xrightarrow[t\to\infty]{} \,\,\frac{\lambda^N}{N!}\, e^{-\lambda}\,.
\end{equation}
From~\eqref{Poisson}, the average number of parked cars is
$\langle N(t)\rangle = \lambda(1-e^{-t})$, which approaches $\lambda$ in the
long-time limit.  

The actual number of parked cars fluctuates about the steady state value
$\langle N\rangle = \lambda$, with the mean deviation from the average equal
to $\sqrt{\langle N^2\rangle -\langle N\rangle^2}=\sqrt{\lambda}$.  Huge
fluctuations are also possible; for example, the parking lot empties with
probability $P_0 = e^{-\lambda}$.  The average time $T$ between successive
lot emptying events roughly scales as the reciprocal of this probability:
$T\sim P_0^{-1} = e^{\lambda}$.  Thus the emptying time is exponentially
large in $\lambda$ for the interesting case of $\lambda\gg 1$; it is
extremely unlikely that the lot is empty when arrival rate of new cars is
large.

One can determine the emptying time by employing the backward Kolmogorov
equation (see, e.g., \cite{R01,KRB10}) that relates the emptying time for a
lot with $n$ cars to the emptying time for a parking lot with $n\pm 1$ cars.
Let $t_n$ be the average time for the lot to empty starting from the state
where $n$ cars are parked.  This emptying time satisfies
\begin{equation}
\label{tn:rec}
t_n = p_n t_{n+1} + q_nt_{n-1} + \delta t_n\,.
\end{equation}
The first term on the right-hand side accounts for the parking of a new car,
an event that occurs with probability $p_n=\lambda/(\lambda +n)$.  After this
event, the average time for the parking lot to empty is $t_{n+1}$.  The
second term accounts a car leaving, which occurs with probability
$q_n=n/(\lambda+n)$.  Finally, $\delta t_n=1/(\lambda +n)$ is the average
time for the number of parked cars to change from $n$ to $n\pm 1$.

Recurrences of the form \eqref{tn:rec} for general $p_n, q_n$ and
$\delta t_n$, are solvable (see, e.g., Chap.\ 12 of Ref.~\cite{KRB10} and
also \cite{KT75}).  Specializing the solution given there to the present
case, the emptying time of the lot when $n$ cars are parked is
\begin{equation}
\label{tn:sol}
t_n=\frac{1}{\lambda}\sum_{j=0}^{n-1}\frac{j!}{\lambda^j}\sum_{i\geq j} \frac{\lambda^i}{i!}\,.
\end{equation}
In the case where a single car is parked, this result simplifies to
$t_1 =(e^\lambda-1)/\lambda$.  In the relevant case of $\lambda\gg 1$, all
$t_n$ with $n\lesssim \lambda$ exhibit this same asymptotic behavior.  For
$\lambda\gg 1$, it is overwhelmingly likely that starting from a lot with a
single parked car, the lot will quickly fill in a time of the order of 1 to
its stationary value of $\lambda$ parked cars.  

\section{Meek Strategy}
\label{sec:meek}

Models that resemble the meek strategy arise in various contexts, e.g., they
have been used to mimic the evolution of genomic DNA~\cite{Li,MLA05a,MLA05b}
and they have been applied to modeling microtubule
dynamics~\cite{AKRMC07,AKR07,LMJ,AES15}.  The meek strategy is actually
identical to the microtubule model discussed in ~\cite{AKRMC07,AKR07}: a car
that parks after the rightmost car corresponds to the addition of a GTP
(guanosine triphosphate) monomer to the microtubule, and the departure of a
car corresponds to the conversion for GTP to GDP (guanosine diphosphate).  A
catastrophe arises when the active end of a microtubule consists of only GDP
monomers; these detach quickly, leading to a rapid decrease in the
microtubule length.  This latter event corresponds to a sudden drop in the
span of parked cars when the rightmost car leaves and the next parked car is
much closer to the target.  The microtubule model is tractable, and the
available analytical results~\cite{AKRMC07,AKR07} provide a rather complete
description of the car distribution that arises in the meek strategy.  Below
we outline some basic results and outline their derivations.

\begin{figure}[ht]
  \centerline{\includegraphics[width=0.45\textwidth]{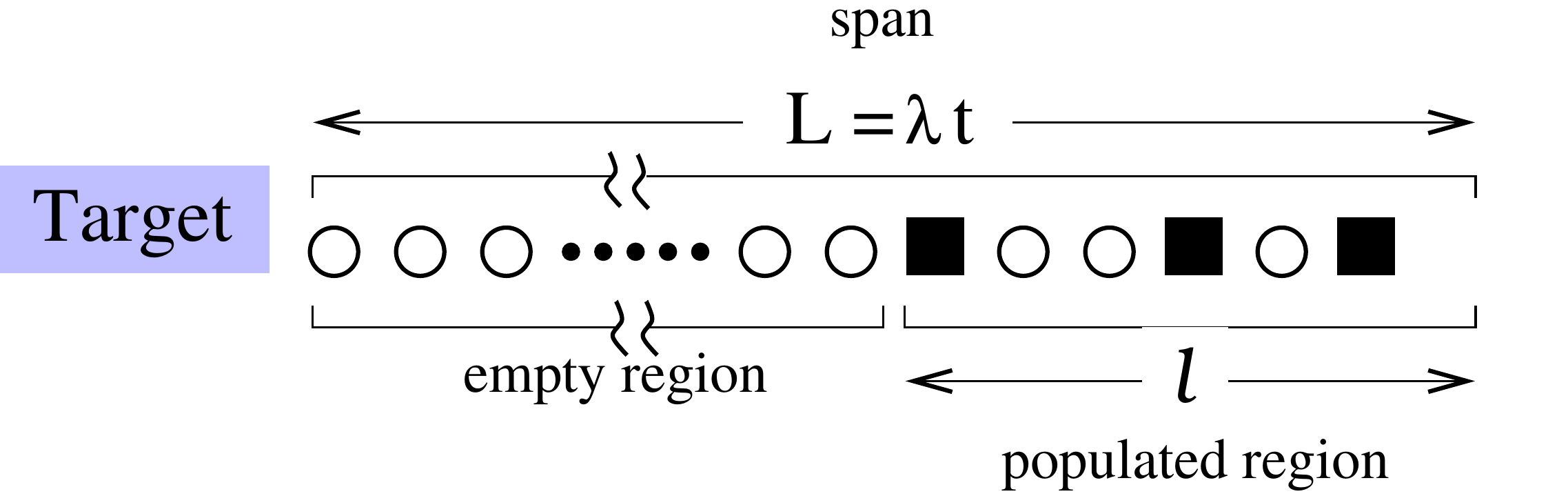}}
  \caption{Schematic of the distribution of parked cars for the meek strategy
    in the large $\lambda$ limit. The span increases linearly in elapsed time
    from the last emptying event. }
\label{populated}  
\end{figure}

\begin{figure}[ht]
  \centerline{\includegraphics[width=0.45\textwidth]{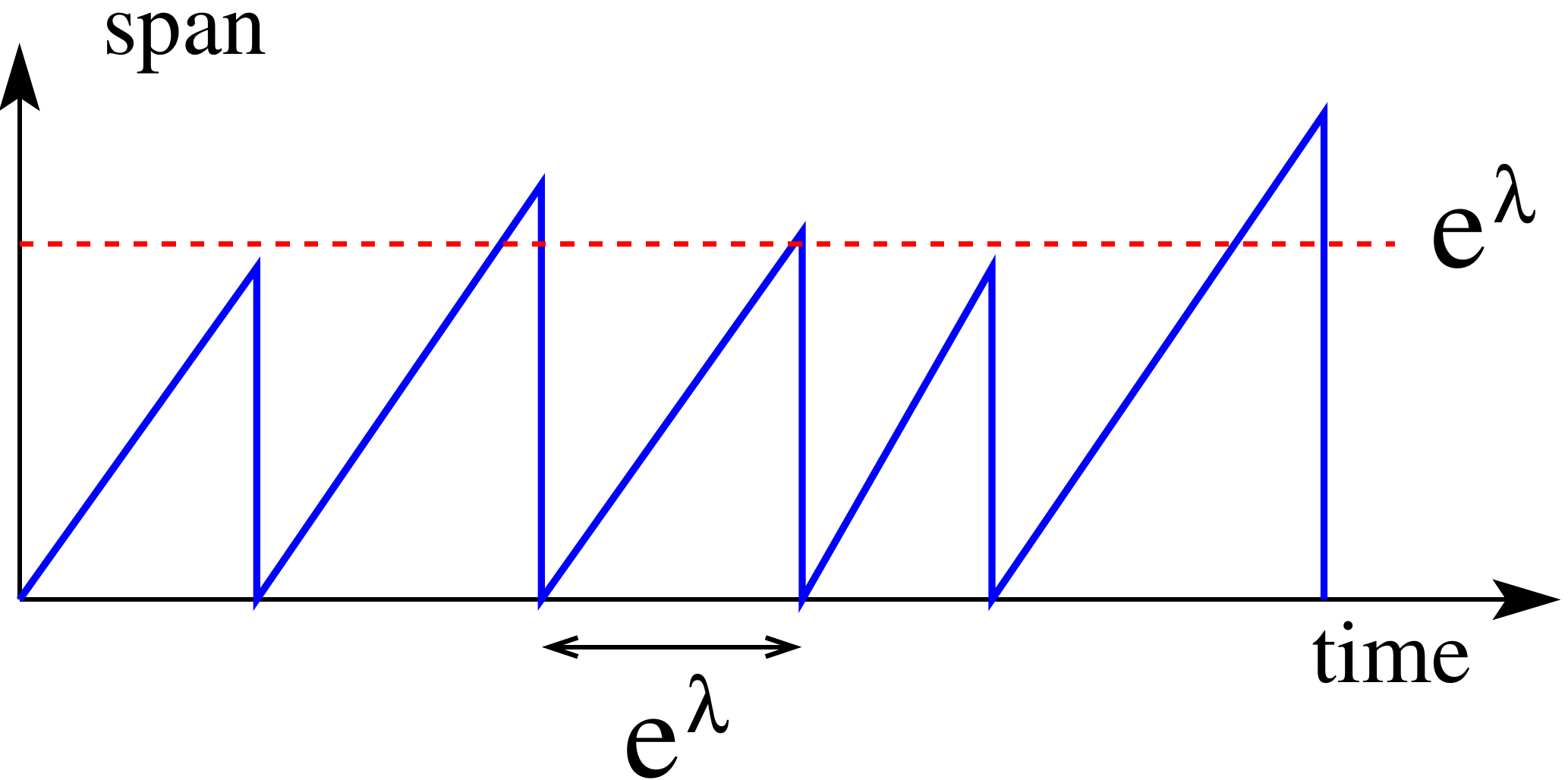}}
  \caption{Schematic picture of the time dependence of the span in the meek
    parking strategy in the large $\lambda$ limit.}
\label{sawtooth}  
\end{figure}

The  meek strategy is ridiculously inefficient
for $\lambda\gg 1$: the typical span, namely the distance from the target to
the rightmost parked car, is huge, $L\propto e^\lambda$, while all cars are
parked within a narrow populated region of length
$\ell\simeq \lambda\ln\lambda$ near the right edge of the span (see
Fig.~\ref{populated}).  At any given moment the populated region moves nearly
systematically to the right at speed $\lambda$ because of the continuous
arrival of cars at this rate.  Since the number of parked cars is roughly
constant and they occupy a fixed-length region $\ell$, there typically is a
huge empty space to the left of the populated region.  This pattern is
disrupted when a rare event occurs in which all cars in the populated region
leave before a new car enters.  When this happens, the parking process begins
anew from an empty lot.  The span therefore has a sawtooth time dependence
(Fig.~\ref{sawtooth}).  When $\lambda\gg 1$, the emptying time is
$T\simeq e^\lambda/\lambda$, and this gives an estimate $L\propto e^\lambda$:
the emptying time is so large that we can only observe the
growth of the span in simulations.

Let us estimate the behavior of the length $\ell$ of the populated region in
the $\lambda\to \infty$ limit.  Since cars leave the lot at rate 1, the
probability that the car that is a distance $x$ from the rightmost car has
not left the lot is $p(x)=e^{-\tau}=e^{-x/\lambda}$.  To estimate the size of
the populated region, we use the fact that the probability that there is a
parked car located a distance $\ell$ or greater from the rightmost car is
\begin{equation}
\label{estimate}
\sum_{x\geq \ell}e^{-x/\lambda}= e^{-\ell/\lambda}/\big(1-e^{-1/\lambda}\big)
\simeq \lambda\, e^{-\ell/\lambda}\,.
\end{equation}
Setting this quantity to 1 gives a simple extreme statistics estimate for the
size of the populated region (see, e.g., Refs.~\cite{G87,G04}).
\begin{equation}
\label{l-av}
\ell=\lambda\ln\lambda\,.
\end{equation}
Thus the length of the populated region is tiny compared to the span.

In the optimistic and prudent strategies the spatial distribution of parked
cars quickly reaches and remains in a quasi-stationary state until the
parking lot empties.  Because newly arriving cars can park in
the interior of the lot, spatial fluctuations in the span are of the order of
$\sqrt{\lambda}$.  Once again, a typical simulation with $\lambda \gg 1$ does
not extend to the emptying time, so all that can be observed is the quasi
steady-state behavior.  We now study the spatial distribution of parked cars
and related features for these two parking strategies.

\section{Optimistic Strategy}
\label{sec:opt}

The key feature of the optimistic strategy is that the dynamics of occupancy
at any spot $i$ depends only on spots $1,2,\ldots,i$; all spots to right can
be ignored. This property occurs in a number of one-dimensional systems that
enjoy {\em spatial causality}, see
e.g. \cite{JML03,JML12,AA-KM,AA11,PLK-JML}, and it allows us to treat the
optimistic strategy analytically.

Denote by $\sigma_j$ the occupation number of spot $j$:
\begin{equation*}
\sigma_j =
\begin{cases}
1 & \text{if $j$ is occupied}\,,\\
0 & \text{if $j$ is empty}\,.
\end{cases}
\end{equation*}
The density $\rho_1=\langle \sigma_1\rangle$ at the first parking spot
satisfies
\begin{equation}
\label{n1-eq}
\frac{d\rho_1}{dt} \equiv \dot \rho_1= \lambda ( 1-\rho_1) -\rho_1\,,
\end{equation}
which simply states that if the first spot is empty, it refills at rate
$\lambda$, while if this spot is occupied, it empties with rate 1.  The
solution to this equation is
\begin{equation}
  \rho_1 = \frac{\lambda}{1+ \lambda}\left[1- e^{-(1+\lambda)t}\right]
\xrightarrow[t\to\infty]{} \,\,   \frac{\lambda}{1+ \lambda}
\end{equation}

Following the same logic as in \eqref{n1-eq}, the density $\rho_k$ satisfies
the equation of motion
\begin{equation}
\label{rk-eq}
\dot \rho_k=\lambda \,\Big\langle (1-\sigma_k)\prod_{j=1}^{k-1}\sigma_j\Big\rangle - \rho_k\,.
\end{equation}
for any $k\geq 2$.  These exact equations are not closed, viz., they involve
the multisite averages $\langle \sigma_1\sigma_2\ldots\sigma_{k-1}\rangle$
and $\langle \sigma_1\sigma_2\ldots\sigma_{k-1}\sigma_k\rangle$.  Writing
evolution equations for these two averages involve other multisite averages.
However, since the occupancy dynamics of the spots $1,2,\ldots,k$ does not
depend on the spots $i>k$, the system of differential equations for $2^k-1$
multisite averages $\langle \sigma_a\ldots\sigma_b\rangle$, with
$1\leq a<\ldots< b\leq k$, is {\em closed}.  While the number of equations
rapidly grows with $k$, they are linear, solvable, and can be treated
recursively.  We now illustrate this approach for $k=2$ and $k=3$.

\subsection{$k=2$}
\label{subsec:2}

When $k=2$, Eq.~\eqref{rk-eq} becomes
\begin{subequations}
\begin{align}
\label{r2-eq}
\dot \rho_2 &=\lambda \rho_1 -\rho_2 - \lambda\langle \sigma_1 \sigma_2\rangle=\lambda \rho_1 -\rho_2 - \lambda \rho_{12}\\
\label{r12-eq}
\dot \rho_{12} &=
\lambda\langle (1-\sigma_1)\sigma_2\rangle + \lambda\langle \sigma_1(1-\sigma_2)\rangle - 2  \rho_{12}
= \lambda (\rho_1 + \rho_2) - 2(1+\lambda)\rho_{12}\,,
\end{align}
\end{subequations}
where $\rho_{12}\equiv \langle \sigma_1 \sigma_2 \rangle$.  The full
time-dependent solutions to these equations are elementary but cumbersome.
In the steady state
\begin{subequations}
\begin{align}
    \label{r1212}
& 0=\lambda \rho_1 -\rho_2 - \lambda\rho_{12}\\
& 0=\lambda (\rho_1 + \rho_2) - 2(1+\lambda)\rho_{12}\,,
\end{align}
\end{subequations}
from which
\begin{align}
\label{r12-sol}
 \rho_{12}  = \frac{\lambda^2}{\lambda^2+ 2\lambda + 2} \qquad\qquad
\frac{\rho_2}{\rho_{12}} = \frac{\lambda+2}{\lambda+1}\,.
\end{align}

\subsection{$k=3$}
\label{subsec:3}

The equation for the density at site 3 is
\begin{equation}
\label{r3-eq}
\dot\rho_3=\lambda (\rho_{12}  - \rho_{123}) -\rho_3\,,
\end{equation}
which involves the three-site average
$\rho_{123}=\langle \sigma_1 \sigma_2 \sigma_3\rangle$.  This average
satisfies
\begin{align}
\label{r123-eq}
\dot \rho_{123} &=\lambda \langle (1-\sigma_1)\sigma_2\sigma_3\rangle
       + \lambda \langle \sigma_1(1-\sigma_2)\sigma_3\rangle
    + \lambda\langle \sigma_1 \sigma_2 (1-\sigma_3)\rangle - 3  \rho_{123}\nonumber\\
     &=    \lambda(\rho_{12} + \rho_{23}+\rho_{13}) - 3(1+\lambda)\rho_{123}\,.
\end{align}
From \eqref{r12-eq} we already know the nearest-neighbor two-site average
$\rho_{12}=\langle \sigma_1 \sigma_2\rangle$, and we also need the equations
for $\rho_{13}=\langle \sigma_1 \sigma_3\rangle$ and
$\rho_{23}=\langle \sigma_2 \sigma_3 \rangle$:
\begin{align}
\label{r13-r23}
    \begin{split}
&\dot\rho_{13} =
\lambda\langle (1\!-\!\sigma_1)\sigma_3\rangle
+ \lambda\langle \sigma_1\sigma_2(1\!-\!\sigma_3)\rangle - 2  \rho_{13}
= \lambda (\rho_3 \!+\! \rho_{12}\!-\!\rho_{123}) - (2\!+\!\lambda)\rho_{13}\\[2mm]
&\dot\rho_{23}=
\lambda\langle \sigma_1(1\!-\!\sigma_2)\sigma_3\rangle
+ \lambda\langle \sigma_1\sigma_2(1\!-\!\sigma_3)\rangle - 2  \rho_{23}
= \lambda (\rho_{13}\! +\! \rho_{12}\!-\!2\rho_{123}) \!-\! 2\rho_{23}
\end{split}
\end{align}

Solving Eqs.~\eqref{r3-eq}--\eqref{r13-r23} in the steady state gives
\begin{subequations}
\begin{align}
    \label{r123-sol}
\rho_{123} &= \frac{\lambda^3}{\lambda^3+3\lambda^2+ 6\lambda + 6}\,,
\end{align}
and in terms of this quantity, the remaining densities can be written as
\begin{align}
  \label{r3-sol}
\frac{\rho_{3}}{\rho_{123}} &= \frac{\lambda^2 + 4 \lambda + 6}{\lambda^2+ 2\lambda + 2}\\
\label{r23-sol}
\frac{\rho_{23}}{\rho_{123}} &=  \frac{\lambda + 3}{\lambda + 2} \\
  \frac{\rho_{13}}{\rho_{123}} &= \frac{(\lambda+1)(\lambda^2
  + 4 \lambda + 6)}{(\lambda+2)(\lambda^2+ 2\lambda + 2)}
\end{align}
\end{subequations}

\subsection{Decoupling approximation}

While the exact results quickly become unwieldy, they greatly simplify in a
decoupling approximation, in which we replace multi-site averages by the
product of corresponding single-site averages (see, e,g, Ref.~\cite{B75} for
a general discussion of such decoupling approaches).  For example, using
$\rho_{12}=\rho_1 \rho_2$ in \eqref{r1212} gives
\begin{align}
\label{r2-MF}
\rho_2^\text{MF} = \frac{\lambda^2}{\lambda^2+ \lambda + 1}\qquad\qquad
\rho_{12}^\text{MF} = \frac{\lambda^3}{(\lambda^2+ \lambda + 1)(\lambda+1)}\,,
\end{align}
where the superscript MF denotes the mean-field densities that arise from the
decoupling approximation.  Similarly using $\rho_{12}=\rho_1 \rho_2$ and
$\rho_{123}=\rho_1 \rho_2 \rho_3$ in \eqref{r123-eq} gives
\begin{equation}
\label{r3-MF}
\rho_3^\text{MF} = \frac{\lambda^4}{\lambda^4+ \lambda^3 + 2\lambda^2+ 2\lambda + 1}\\
\end{equation}

\begin{figure}
\centering
\subfigure[]{\includegraphics[width=0.4\textwidth]{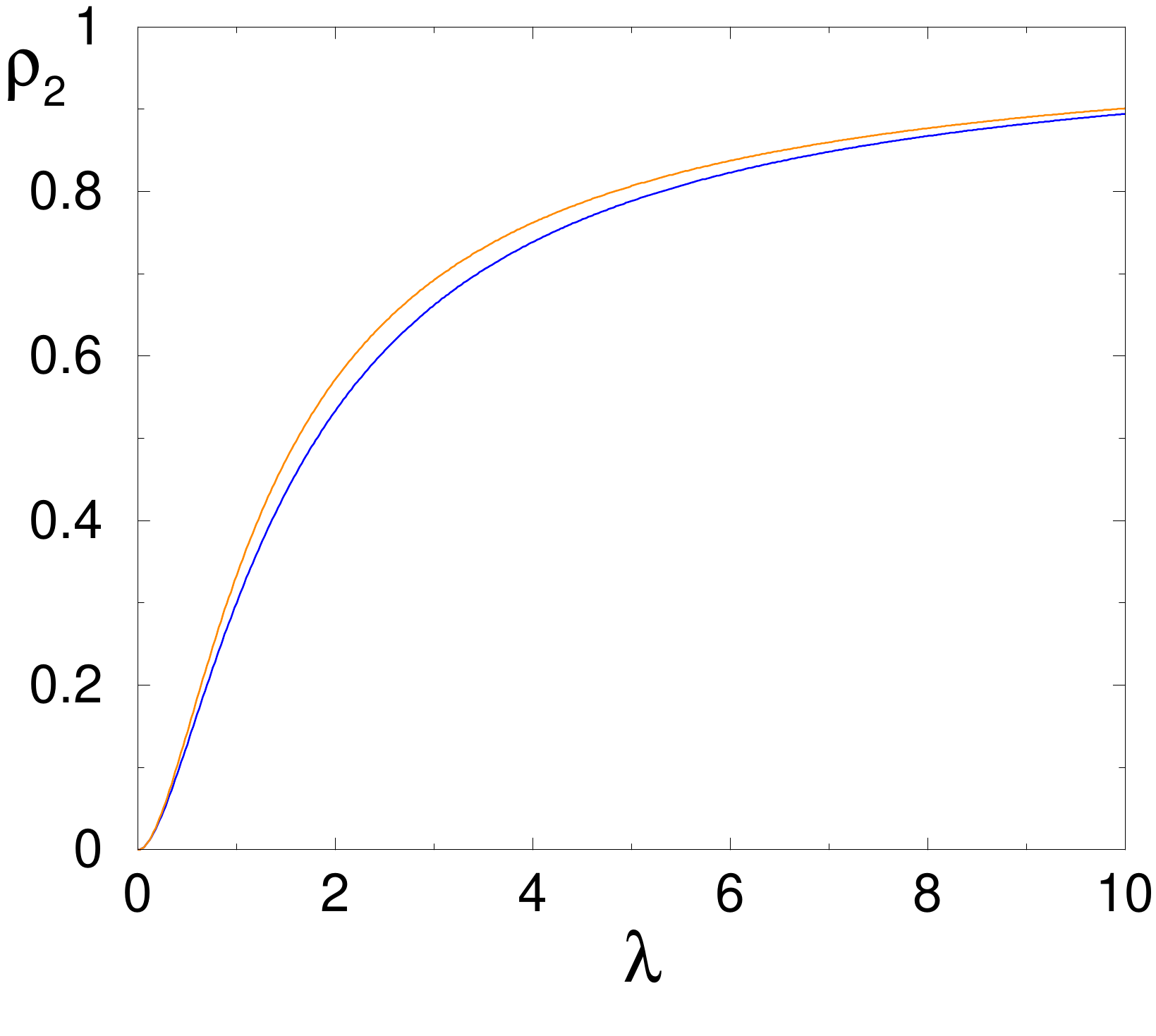}}\qquad
\subfigure[]{\includegraphics[width=0.4\textwidth]{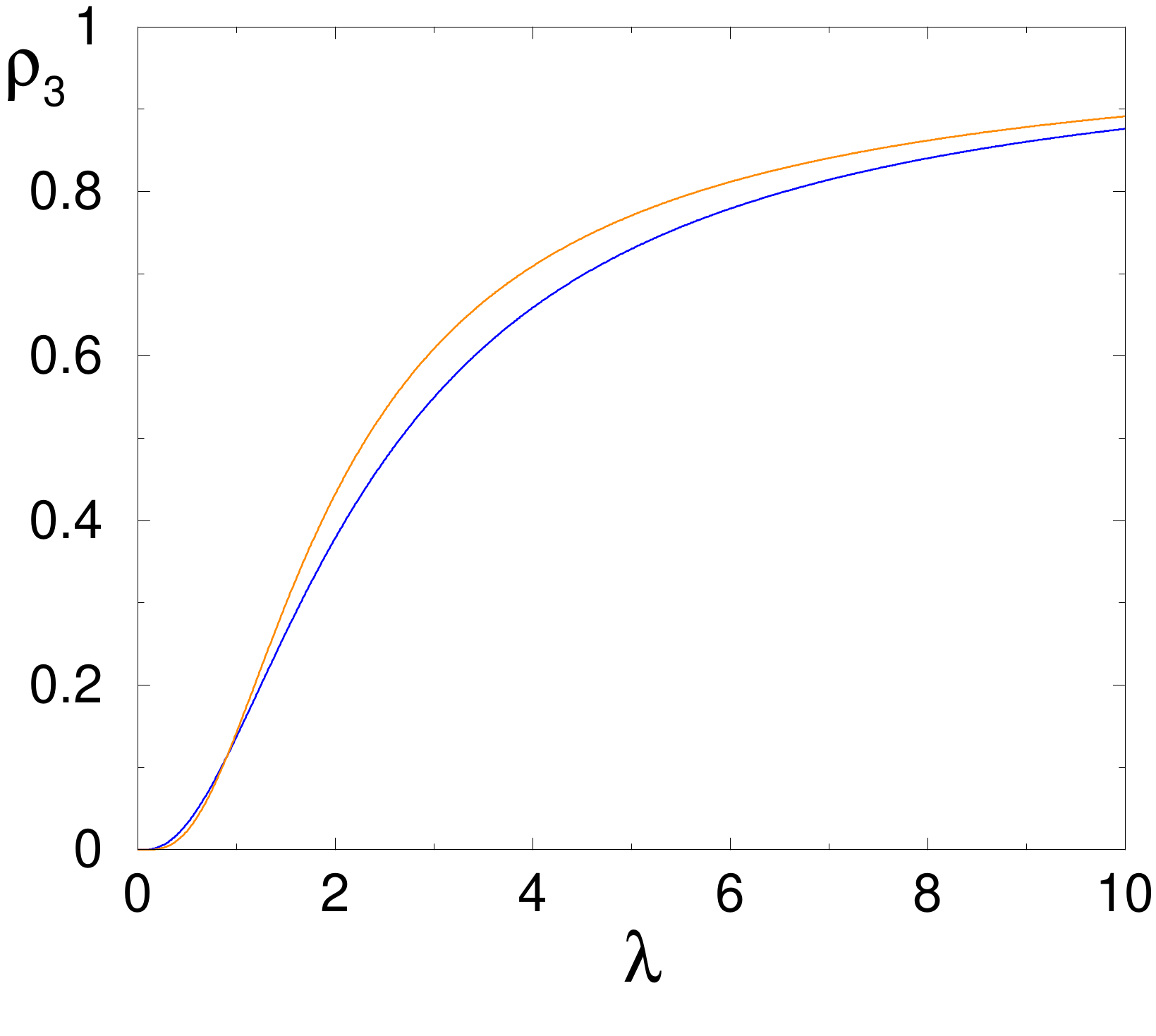}}
\caption{(a) The exact (blue) and decoupling approximation density (orange)
  at site 2 from \eqref{r12-sol} and \eqref{r2-MF}, respectively.  (b) Same
  for site 3. Here the exact density from~\eqref{r123-sol} lies below the
  decoupling approximation \eqref{rk-rec} when $\lambda>0.900966$. }
\label{Fig:r2-MF-exact}
\end{figure}


To gauge the accuracy of the decoupling approximation we compare the exact
and approximate densities $\rho_2(\lambda)$ and $\rho_3(\lambda)$ in
Fig.~\ref{Fig:r2-MF-exact}.  The decoupling approximation is generally
accurate and becomes more so accurate as $\lambda\to \infty$.  To show this
analytically, we define $\epsilon\equiv 1/(\lambda+1)$ as a small parameter,
so that $\rho_1=1-\epsilon$.  The expansion in $\epsilon$ yields
\begin{align*}
\rho_2 &= 1-\epsilon  - 2\epsilon^2 + 2\epsilon^3+2\epsilon^4+\ldots \\
\rho_2^\text{MF} & = 1-\epsilon - \epsilon^2 + \epsilon^4  +\ldots
\end{align*}
for the density and
\begin{align*}
\rho_{12} &= 1-2\epsilon + 2 \epsilon^3 - 2\epsilon^5 + \ldots\\
\rho_{12}^\text{MF} & = 1-2\epsilon + \epsilon^3+\epsilon^4+\ldots
\end{align*}
for the two-site correlation function.  For $\rho_2$, the decoupling
approximation expression is exact to second order in $\epsilon$, and
$\rho_{12}$ is exact to third order.  This pattern seems to hold for
different sites; e.g., for the density at site 3 we expand \eqref{r123-sol}
and \eqref{r3-MF} and find that two leading orders of the expansion are again
exact:
\begin{equation*}
\begin{split}
\rho_3 &= 1-\epsilon  - 4\epsilon^2 - 2\epsilon^3 + \ldots\\
\rho_3^\text{MF} & = 1-\epsilon  - 2\epsilon^2  - 2\epsilon^3 + \ldots
\end{split}
\end{equation*}

\subsection{Large-$k$ behavior}

Because the multisite correlation functions are cumbersome and the decoupling
approximation is accurate, and even asymptotically exact in the most
interesting $\lambda\to\infty$ limit, we now focus on the large-$k$ behavior
using the decoupling approximation.  In the steady state we obtain
\begin{subequations}
\begin{equation}
\rho_{k+1} = \frac{\lambda\prod_{1\leq j\leq k}\rho_j}{1+\lambda\prod_{1\leq j\leq k}\rho_j}\,,
\end{equation}
which can be simplified to
\begin{equation}
\label{rk-rec}
\rho_{k+1} = \frac{\rho_k^2}{1-\rho_k+\rho_k^2}\,.
\end{equation}
\end{subequations}
Starting from $\rho_1 = {\lambda}/(1+ \lambda)$ and iterating \eqref{rk-rec},
we find
\begin{equation}
\label{nk-rec}
n_{k+1} - n_k = \epsilon n_k^2\left[1-\frac{\epsilon^2 n_k^2}{1-\epsilon n_k+\epsilon^2 n_k^2}\right]\,,
\end{equation}
where $\epsilon=1/(\lambda+1)$ and we have written the solution in the form
$1-\rho_k = \epsilon n_k$. 

\begin{figure}[ht]
\begin{center}
\subfigure[]{\includegraphics[width=0.475\textwidth]{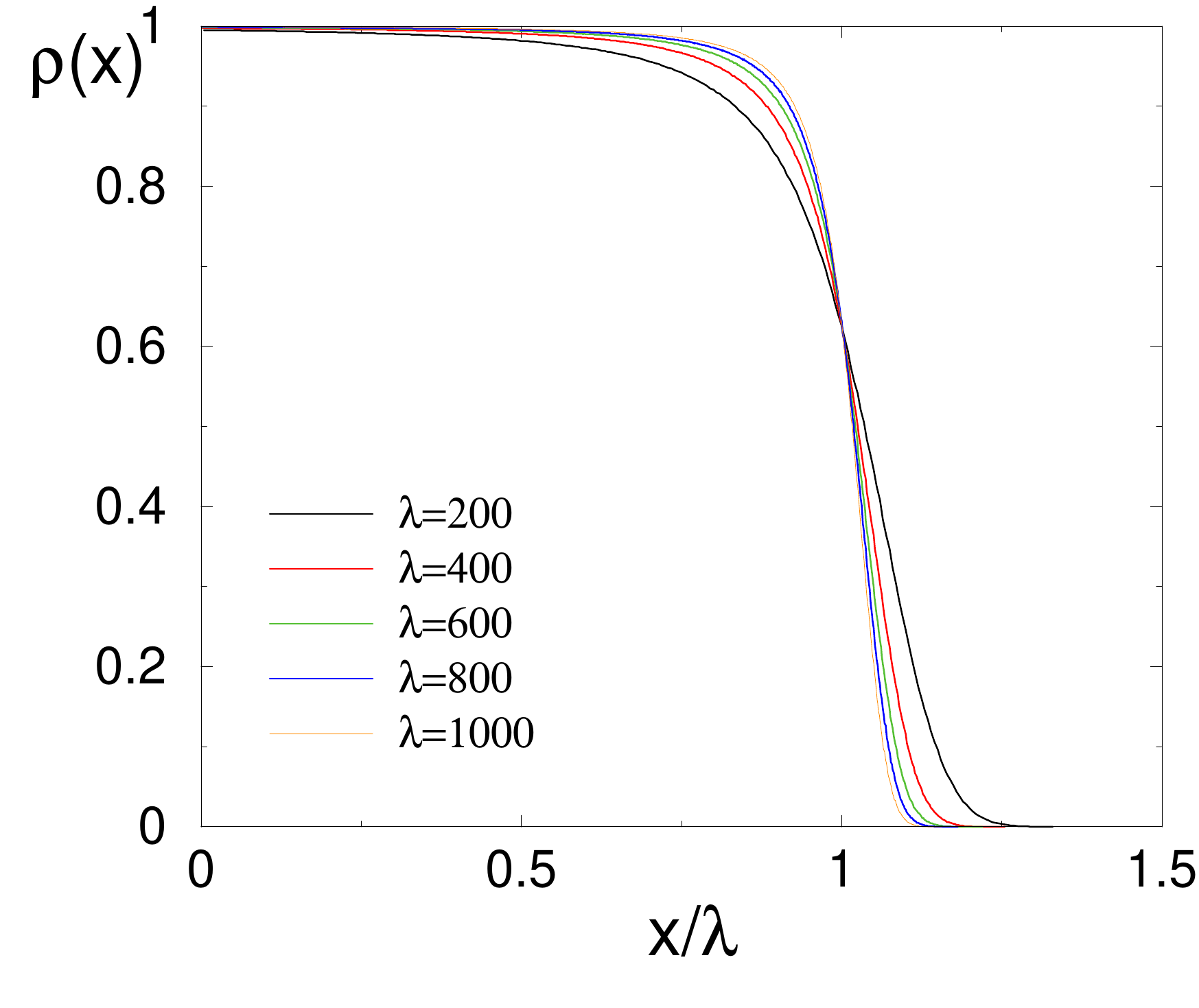}}
\subfigure[]{\includegraphics[width=0.465\textwidth]{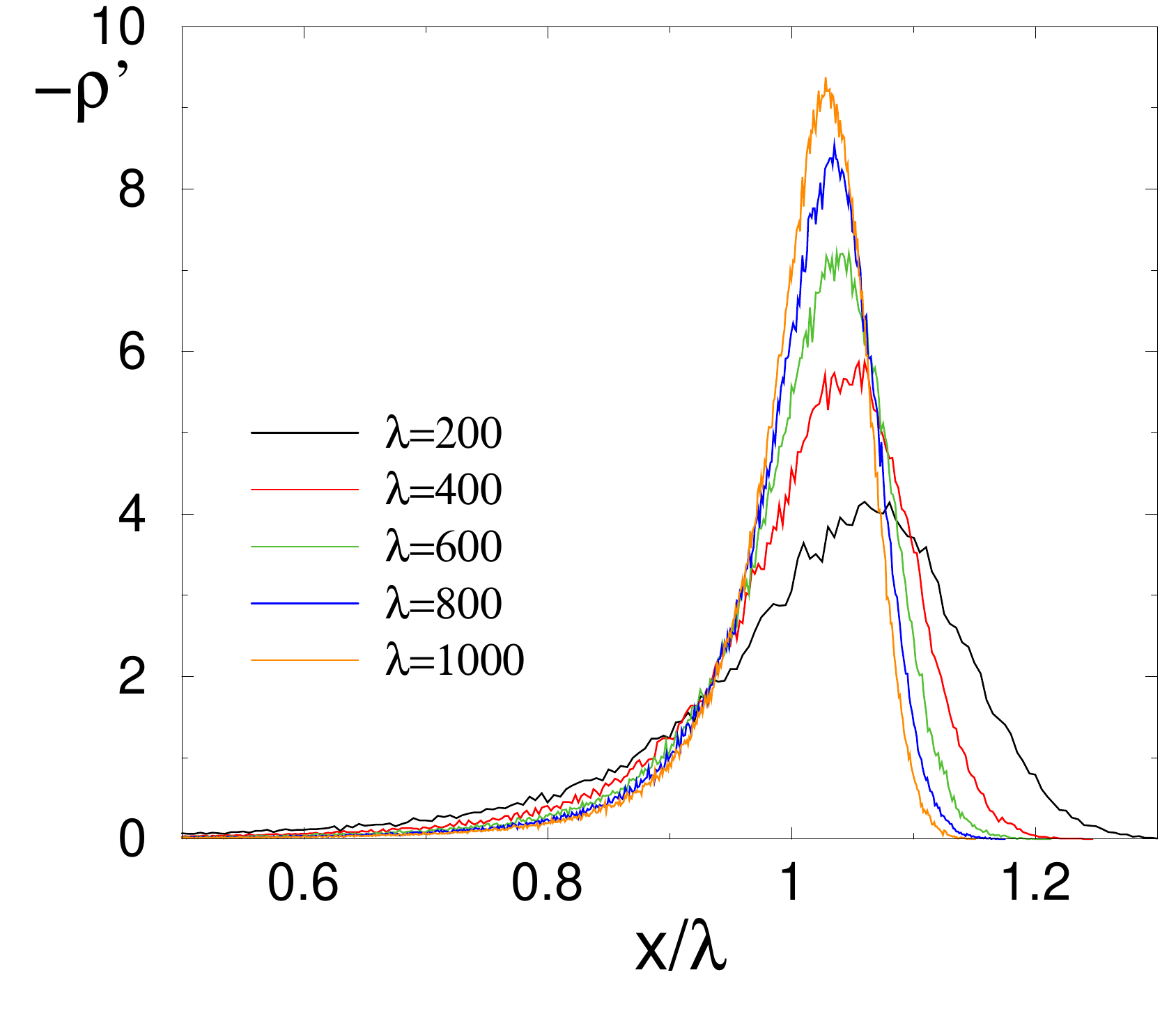}}
\end{center}
\caption{(a) The density of parked cars $\rho(x)$ as a function of distance
  $x$ from the target for the optimistic parking strategy for $\lambda$
  between 200 and 1000.  (b) The derivative $-d\rho/dx$.  These curves have
  been smoothed over 1\% of the data range. }
\label{rho-opt}
\end{figure}

Keeping only the leading term and replacing the difference by the derivative
gives
\begin{equation*}
\frac{d n_k}{dk} = \epsilon n_k^2\,,
\end{equation*}
whose solution, subject to the boundary condition $n_1=1$, yields
\begin{subequations}
\begin{equation}
\label{nk-sol}
n_k = \frac{1}{1-\epsilon (k-1)} =  \frac{\lambda+1}{\lambda+2-k}
\end{equation}
or equivalently
\begin{equation}
\label{rk-sol}
\rho_k = 1 - \frac{1}{\lambda+2-k}
\end{equation}
\end{subequations}
This solution applies in the ``bulk'' region where $ \lambda-k\gg 1$.  For small
$k$, the density remains close to 1, but with a deviation that slows grows as
$k$ increases.

Figure~\ref{rho-opt}(a) shows simulation results for the steady-state density
of parked cars for the optimistic strategy for representative value of
$\lambda$.  Our simulations start with an empty system and continue until
roughly $10^5\lambda$ cars have parked.  To give a more quantitative sense of
the accuracy of \eqref{rk-sol}, Table~\ref{tab:table1} compares the prediction of
this equation with simulation results.

\begin{table}[h!]
\label{tab:table1}
  \begin{center}
    \caption{Comparison of $\rho_k$ from simulations (second column) and from
      Eq.~\eqref{rk-sol} (third column) for the case $\lambda=1000$.}
    \begin{tabular}{r|l|l}
      $k$ & $\rho_k$ (sim.) & $\rho_k$ \eqref{rk-sol}\\
      \hline
      1 & 0.999004 & 0.999001\\
      10 & 0.99898 & 0.99899\\
      100 & 0.99877 & 0.99889\\
      200 & 0.99845 & 0.99875\\
      400 & 0.99724 & 0.99834\\
    \end{tabular}
  \end{center}
\end{table}

As $\lambda$ increases, the density of parked cars becomes more step-like and
resembles the Fermi-Dirac distribution.  To characterize the region near
$x=\lambda$, Fig.~\ref{rho-opt}(b) shows the derivative $-d\rho/dx$.  The
increasing steepness of the density step at $x=\lambda$ in
Fig.~\ref{rho-opt}(a) corresponds to the sharpening of the peak in
Fig.~\ref{rho-opt}(b).  From the latter data, we also measure its width and
find that this width shrinks roughly as $\lambda^{-1/2}$.

\subsection{Vacancies}

Let us now determine the location of the nearest open parking spot, or
vacancy.  The probability $V_k$ that the first vacancy is located at site $k$
is
\begin{subequations}
 \begin{equation}
V_k = \left\langle \prod_{j=1}^{k-1}\sigma_j\,(1-\sigma_k)\!\right\rangle\,,
\end{equation}
which becomes, in the decoupling approximation,
\begin{equation}
\label{Vk}
V_k = \epsilon n_k \prod_{j=1}^{k-1}\rho_j\,.
\end{equation}
\end{subequations}
Taking the logarithm, replacing the sum by integration, and using
\eqref{rk-sol} we get
\begin{align*}
\sum_{j=1}^{k-1} \ln\rho_j \simeq  \int_{1}^{k} dj\,\ln\!\left(1 - \frac{1}{\lambda+2-k}\right) 
\simeq  -\int_{1}^{k}  \frac{dj}{\lambda+2-j} = \ln\frac{\lambda+2-k}{\lambda+1}\,.
\end{align*}
Comparing with \eqref{nk-sol}, the product is
\begin{equation}
\label{prod:nk}
\prod_{j=1}^{k-1}\rho_j \simeq \frac{1}{n_k}\,.
\end{equation}
Thus the location of the first vacancy is uniformly distributed in the range
$[0,\lambda]$:
\begin{equation}
\label{Vk-sol}
V_k = 
\begin{cases}
\epsilon & k<\lambda\\
0           & k > \lambda\,,
\end{cases}
\end{equation}
from which the average position of the first vacancy is
\begin{equation}
\label{vac-av}
v_1\equiv \langle k\rangle = \sum_{k=1}^{\lambda} k V_k = \frac{\lambda}{2}\,.
\end{equation}
Our simulations are in excellent agreement with this simple result. The same
approach can be applied to compute the joint probability $V_{k_1,\ldots,k_m}$
to have $m$ vacancies at sites $k_1<k_2<\ldots<k_m<\lambda$.  This
probability is proportional to the product of the densities at the positions
of all but the last vacancy, viz.,
\begin{equation}
\label{Vkn-sol}
V_{k_1,\ldots,k_m} = \epsilon^m \prod_{a=1}^{m-1} n_{k_a}\,.
\end{equation}
with $n_k$ given by \eqref{nk-sol}.  In general, the
average position of the $m^\text{th}$ vacancy is
\begin{equation}
\label{vn}
v_m = (1-2^{-m})\lambda\,.
\end{equation}
We anticipate that this result will hold as long as the $m^\text{th}$ vacancy
is in the bulk of the density distribution.

\section{Prudent Strategy and Comparison to Optimistic Strategy}
\label{sec:prudent}

\begin{figure}[ht]
\begin{center}
\subfigure[]{\includegraphics[width=0.45\textwidth]{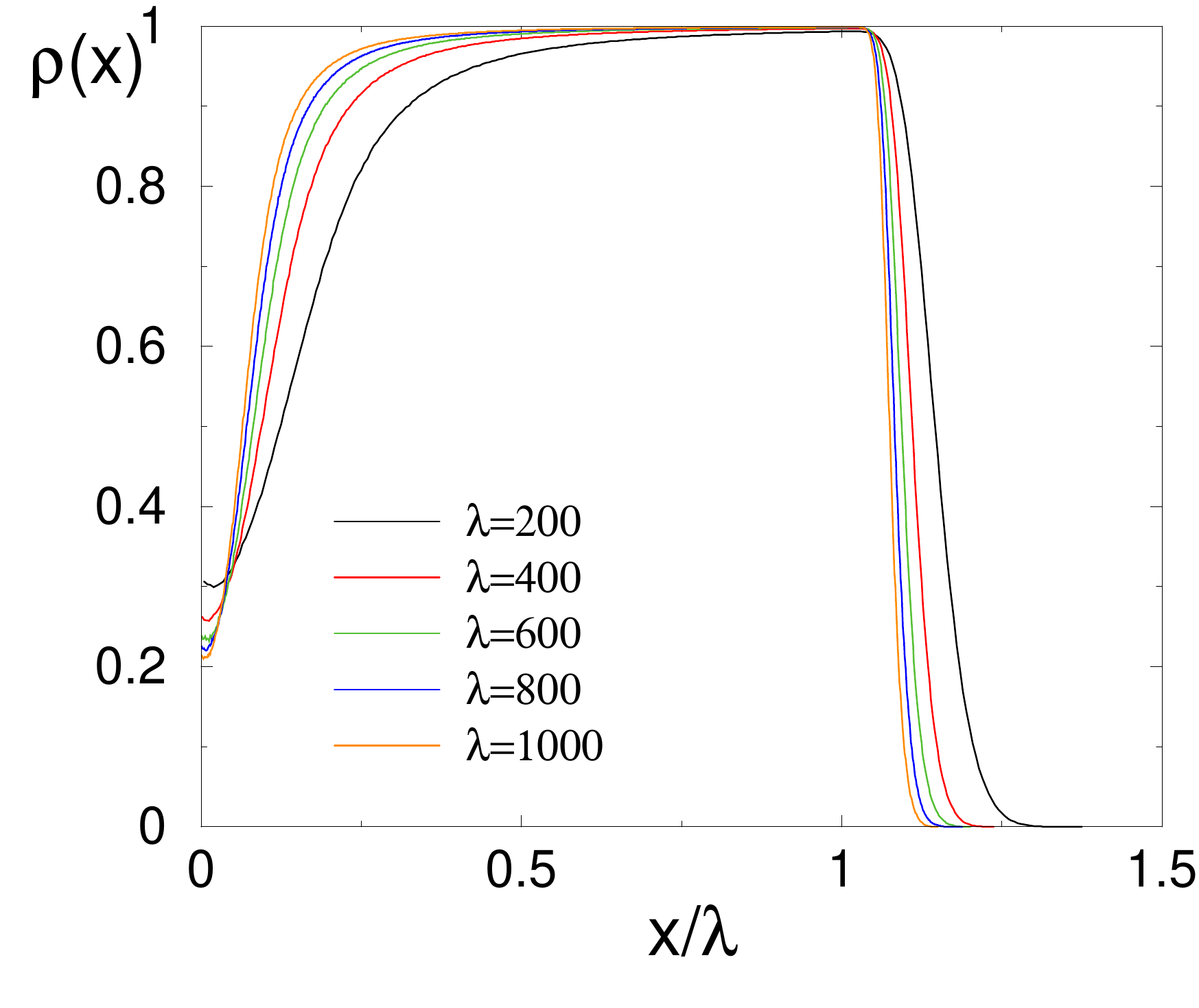}}\qquad
\subfigure[]{\includegraphics[width=0.425\textwidth]{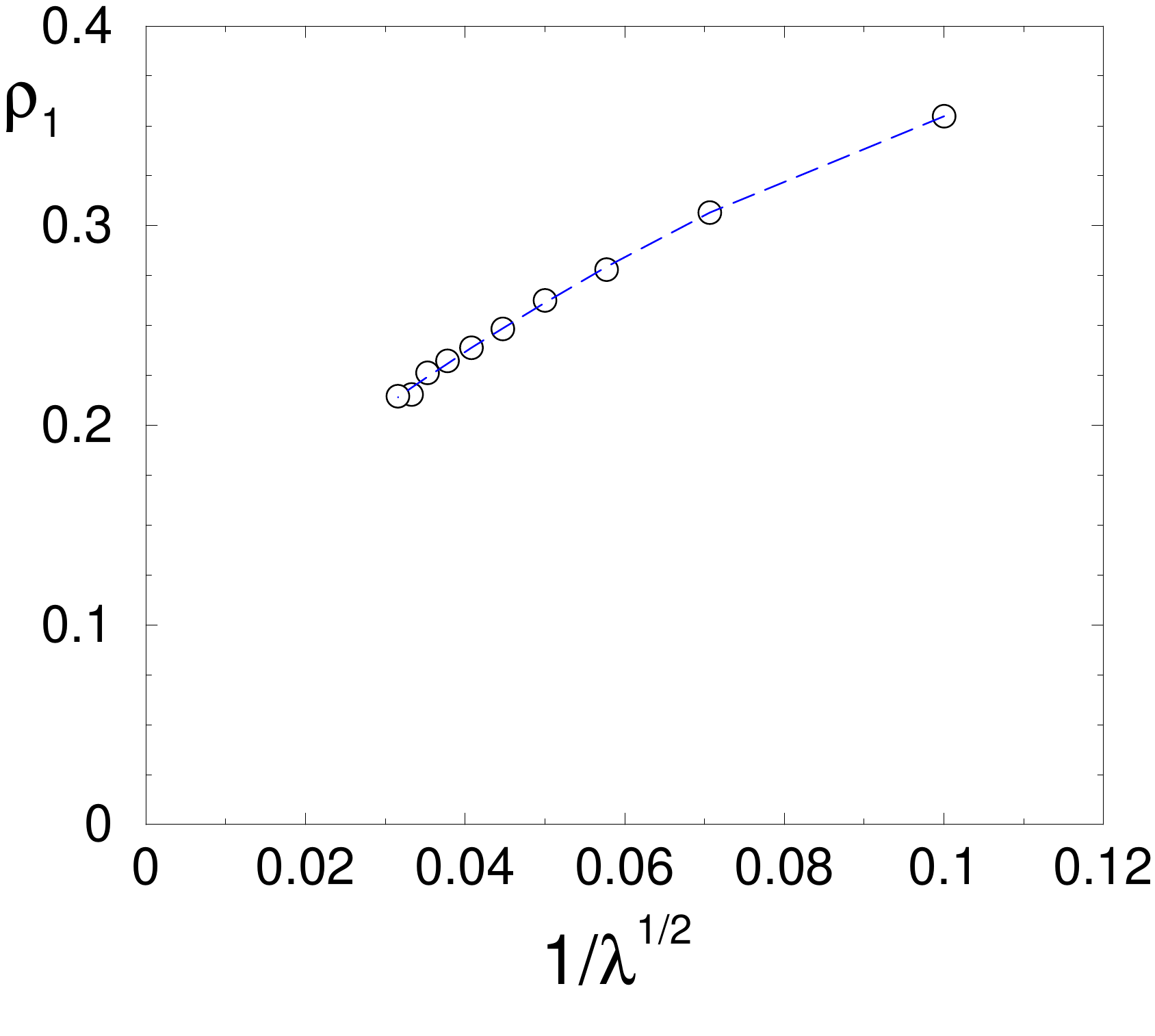}}
\end{center}
\caption{(a) The density of parked cars $\rho(x)$ as a function of distance
  $x$ from the target for the prudent parking strategy for $\lambda$ between
  200 and 1000.  (b) The average density of parked cars at the closest
  parking spot to the target, $\rho_1(\lambda)$, as a function of
  $1/\lambda^{1/2}$ for $\lambda$ in the range between 100 and 1000.  The
  dashed line is a quadratic fit to these data that extrapolates to
  $\rho_1(\infty)\approx 0.11$.}
\label{rho-rho1}  
\end{figure}

The many-body nature of the parking process is more complicated for the case
of the prudent strategy and our results for this case are simulational.
Figure~\ref{rho-rho1}(a) shows the steady-state density of parked cars for
the prudent strategy for representative value of $\lambda$.  The salient
feature of the prudent strategy is that there are lots of open parking spots
very close to the target.  This feature arises because a newly arriving car
only penetrates to the first vacancy (or contiguous vacancy cluster) that it
encounters.  Thus parking spots that are close to the target are ``screened''
by more distant spots.  Because it is unlikely that a new car penetrates to
close parking spots and these spots open up with rate 1, the density of open
spots near the target are likely to be high.

To check this last hypotheses, we plot simulation data for the density of
parked cars at site 1 as a function of $\lambda$ (Figure~\ref{rho-rho1}(b)).
The data indicate that the average density of parked cars at the first spot,
$\rho_1(\lambda)$, is a systematically decreasing function of $\lambda$ that
extrapolates to a non-zero value for $\lambda\to\infty$.  The quadratic fit
shown in this figure extrapolates to $\rho_1(\infty)\approx 0.11$.

\begin{figure}[ht]
\begin{center}
\subfigure[]{\includegraphics[width=0.45\textwidth]{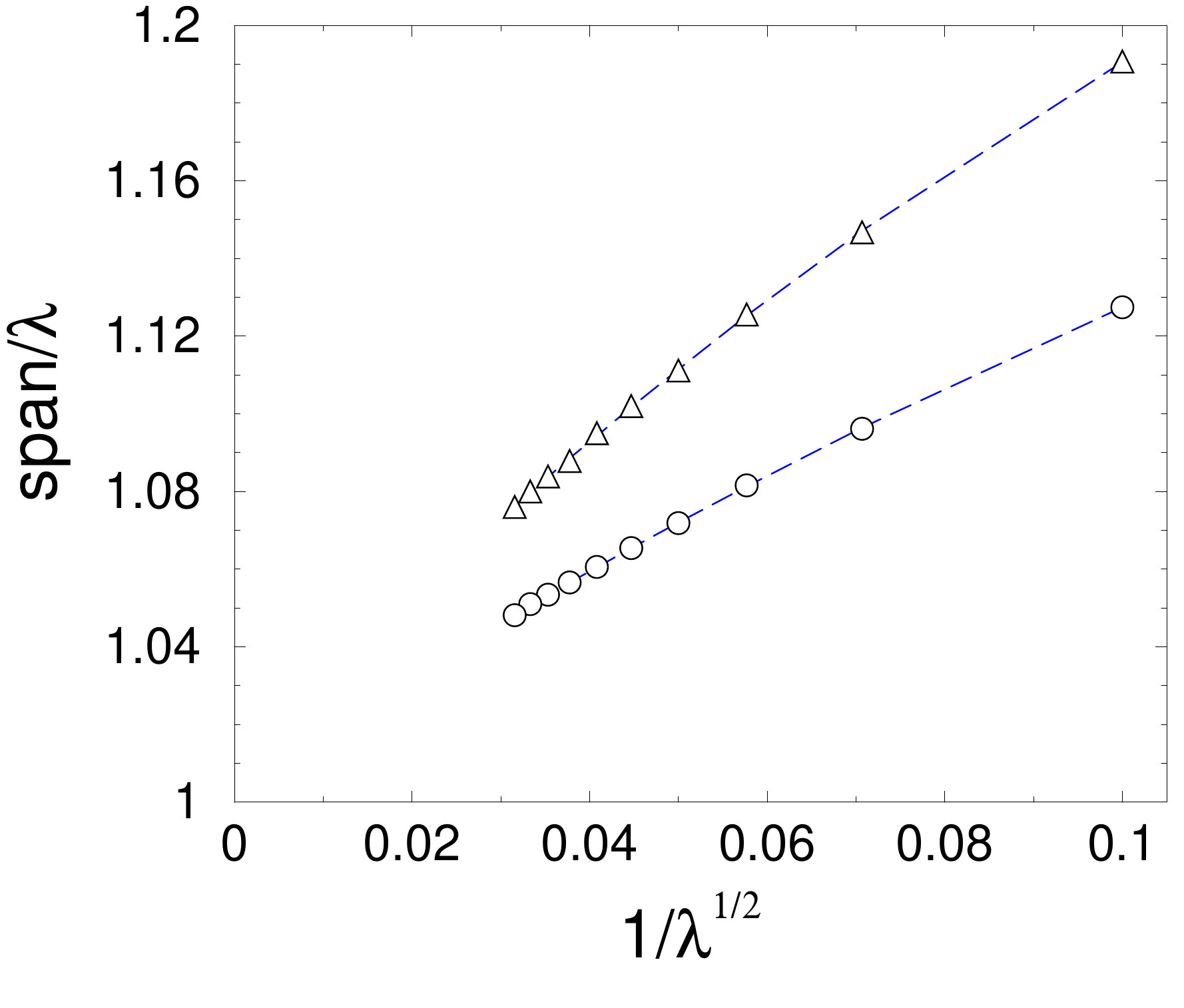}}
\subfigure[]{\includegraphics[width=0.435\textwidth]{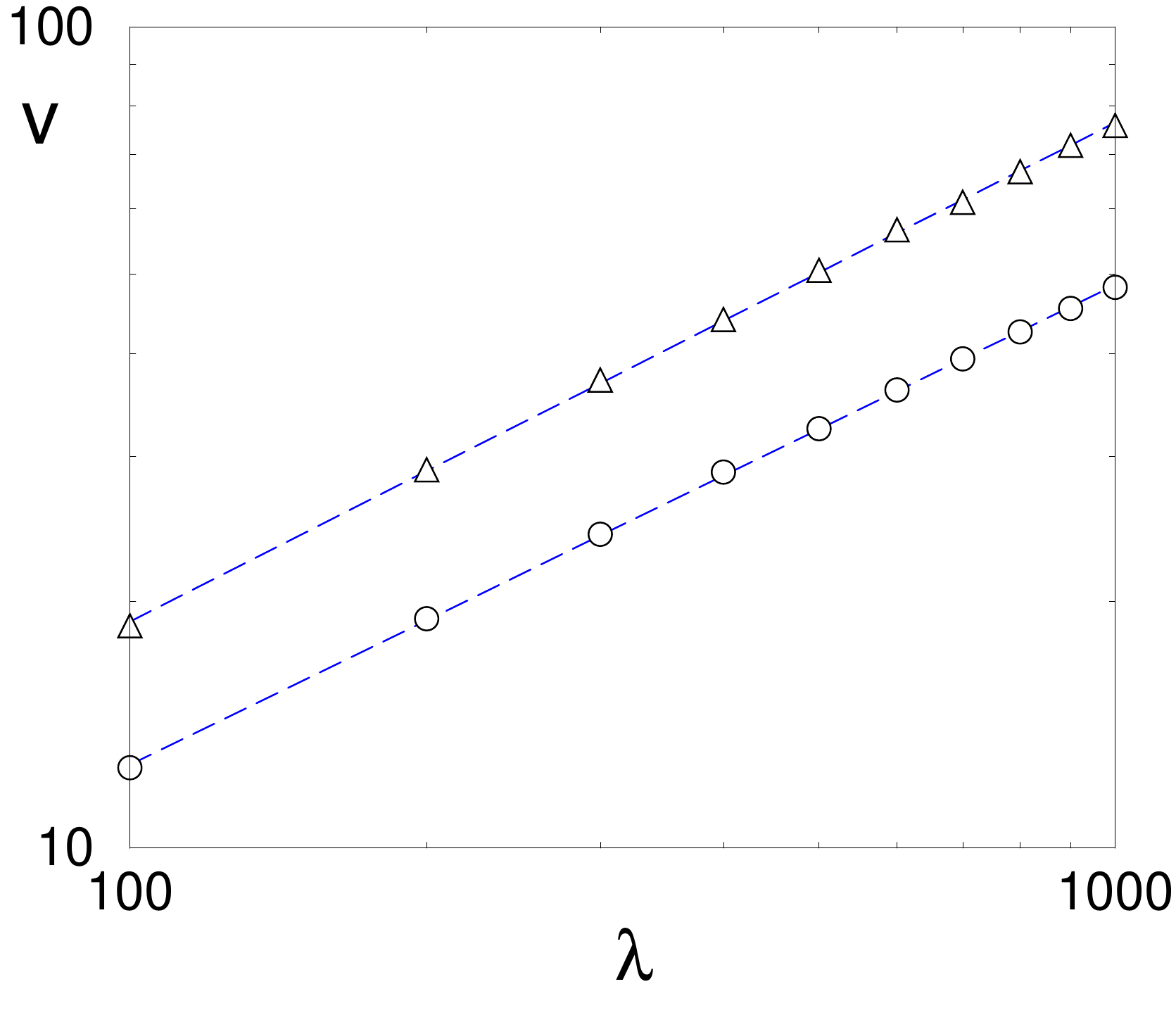}}
\end{center}
\caption{(a) The average span for the optimistic ($\circ$) and prudent
  ($\Delta$) strategies divided by $\lambda$ as a function of
  $1/\lambda^{1/2}$ for $\lambda$ in the range between 100 and 1000.  The
  dashed curves are quadratic fits to these data.  (b) The average number of
  open parking spots as a function of $\lambda$ for the optimistic ($\circ$)
  and prudent strategies ($\Delta$).}
\label{span-nv}  
\end{figure}

For both the optimistic and prudent strategies, the average span $L$ appears
to have the asymptotic behavior $L\simeq \lambda + a\lambda^{1/2}$; more precisely
\begin{equation}
\label{L-a}
\lim_{\lambda\to\infty} \frac{L-\lambda}{\sqrt{\lambda}}=a
\end{equation}
The amplitude $a>0$ is larger for the prudent strategy.  Equation \eqref{L-a} 
implies that the number of vacancies grows as $a \sqrt{\lambda}$.  Thus both
the optimistic and prudent strategies are efficient in that there are
generally very few open parking spots in the steady state.  Figure~\ref{span-nv}(a)
shows $L/\lambda$ plotted versus $1/\lambda^{1/2}$ for both the optimistic
and prudent strategies.  Both datasets show the same qualitative behavior in
which $L/\lambda$ appears to extrapolate to 1 for $\lambda\to\infty$, with
corrections that vanish as $1/\lambda^{1/2}$.

Figure~\ref{span-nv}(b) shows the dependence of the number of open parking spots on
$\lambda$ for the both optimistic and prudent strategies.  In both cases, the
number of parking spots, $v$, appears to grow as $\lambda^\nu$, with
$\nu\approx 0.58$.  However, these data have a slight downward curvature and
we expect that asymptotically $v\sim \lambda^{1/2}$.  

\begin{figure}[ht]
\begin{center}
\includegraphics[width=0.435\textwidth]{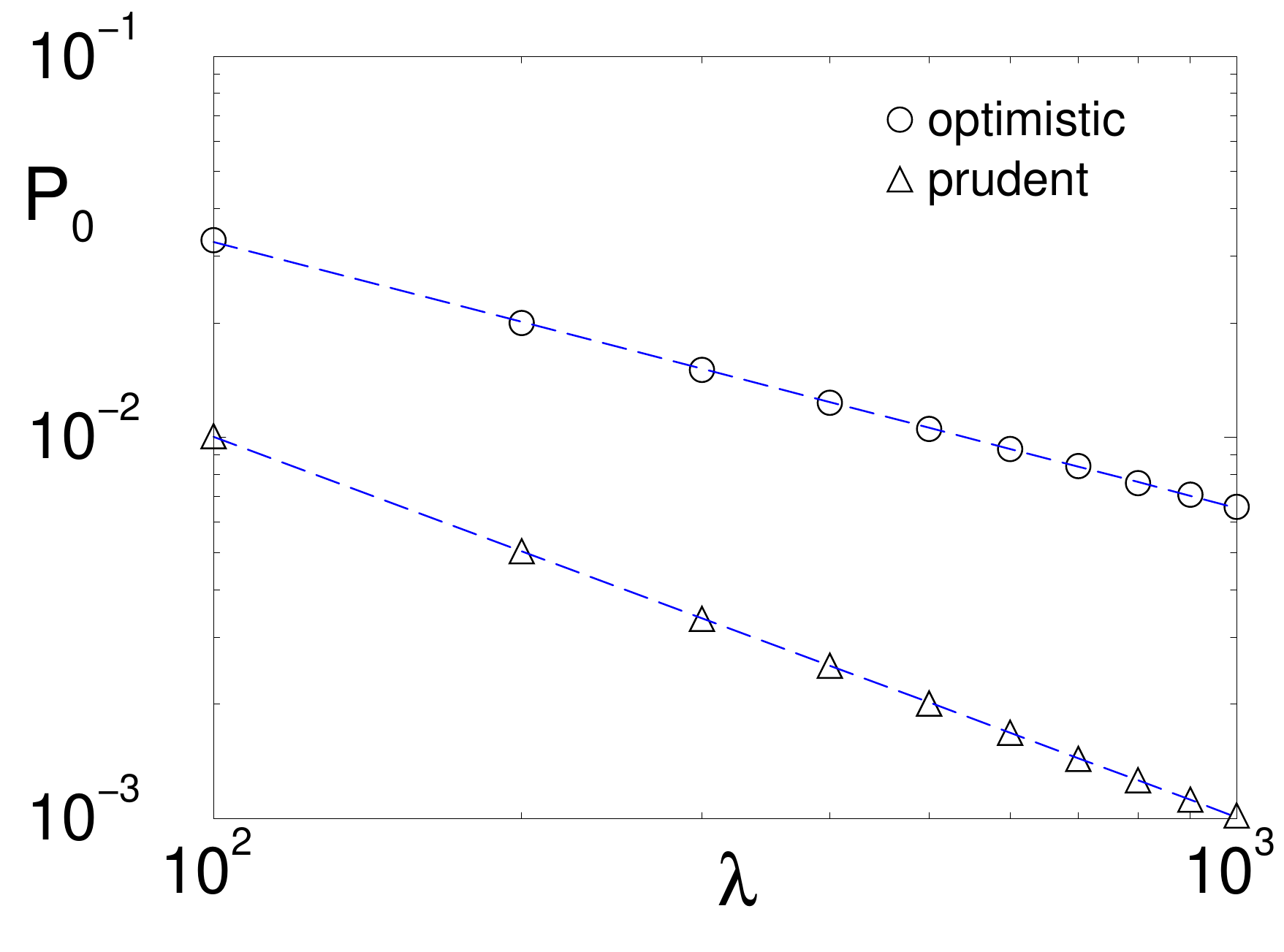}
\end{center}
\caption{The Probability $P_0$ that the parking lot contains no open
  parking spaces as a function of $\lambda$.}
\label{P0}  
\end{figure}

A related measure of parking efficiency is the fraction of times that a
driver has to backtrack to the end of the parked cars because there are no
open spots available.  This is the same as the probability $P_0$ that there
are no open parking spots in the lot.  As shown in Fig.~\ref{P0}, the
fraction of parking attempts that requires the driver to backtrack to the end
of the parking lot varies as $\lambda^{-\alpha}$ for both strategies, with
$\alpha\approx1$ for the prudent strategy and $\alpha\approx 0.7$ for the
optimistic strategy.  

\section{Discussion}

We introduced simple parking strategies in an idealized one-dimensional
parking lot, viz. the semi-infinite line, with cars arriving one at a time at
rate $\lambda$ and departing at rate 1.  We assume that successive car
arrivals are sufficiently separated in time that there is no competition
between cars trying to park in the same spot.  The number of parked cars is
independent of the parking strategy.  This number obeys a Poisson
distribution, with the average number of parked cars equal to $\lambda$.
However, the spatial distribution of parked cars strongly depends on the
strategy that is employed.

In the meek strategy, each new car parks behind the most distant parked car.
While this might be a reasonable approach when $\lambda$ is small, it quickly
becomes ludicrous for large $\lambda$ because the position of the last car is
typically a distance of the order of $e^\lambda$ from the target.  However,
if there are a few meek drivers while the majority follow the prudent or
optimistic strategy, then the meek strategy is not bad because meek drivers
will park a distance $\lambda$ from the target.

Much more practical are the optimistic and prudent strategies.  In the
optimistic strategy, a driver hopes that there is parking spot close to the
target.  Thus the driver goes all the way to the target, ignores all open
spots, and finally parks at the first spot encountered upon backtracking.  In
the prudent strategy, the driver does not have the same degree of confidence
but hopes that an open spots exists that is closer to the target than the
most distant parked car; specifically, the prudent driver parks at the left end 
of the first gap. 

\begin{figure}[ht]
\begin{center}
\subfigure[]{\includegraphics[width=0.425\textwidth]{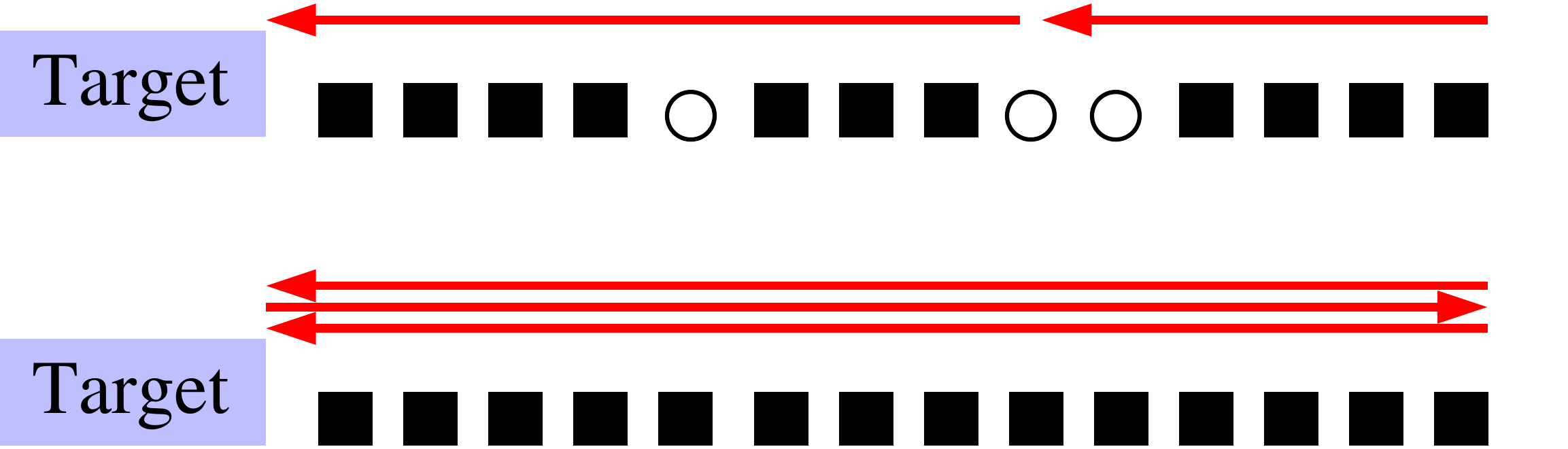}}\qquad
\subfigure[]{\includegraphics[width=0.425\textwidth]{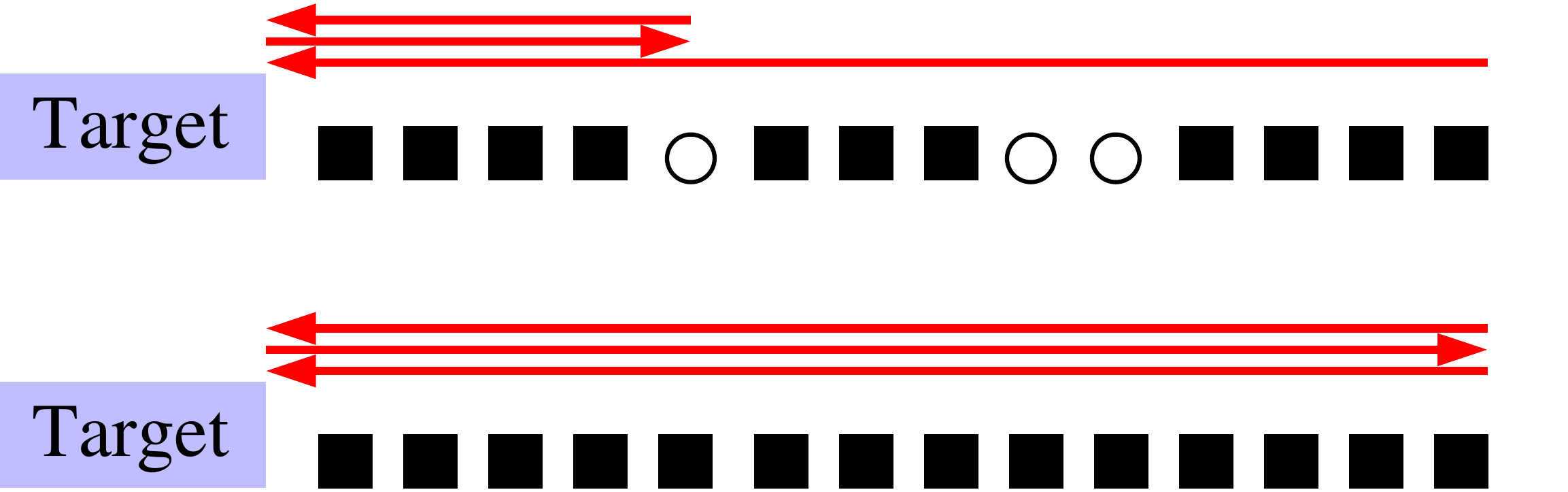}}
\end{center}
\caption{Schematic definition of the parking cost for the (a) prudent and
  (b) optimistic strategies for a parking lot with vacancies (top line) and
  for a full lot (bottom line).}
\label{cost}  
\end{figure}

Which strategy---optimistic or prudent---is better?  To address this question
quantitatively, one must introduce a cost of a parking event and compare the
costs of the two strategies.  A natural definition of cost is the distance
from the parking spot to the target plus the time wasted in looking for a
parking spot.  To minimize the number of parameters, we assume that the speed
of the car in the parking lot is the same as the walking speed.  Thus an
appropriate cost measure is the distance traveled by the car in the lot plus
the distance that the driver walks from the parking spot to the target
(Figure~\ref{cost}).  With this definition of parking cost, the average cost
scales linearly with $\lambda$, but with different prefactors for the
optimistic and prudent strategies.  On average, the prudent strategy is less
costly.  Thus even though the prudent strategy does not allow the driver to
take advantage of the presence of many prime parking spots close to the
target, the backtracking that must always occur in the optimistic strategy
outweighs the benefit by typically parking closer to the target.

Needless to say, there are other ways to judge the efficacy of a parking
strategy.  Psychologically, a prudent driver may get upset by parking far
from the target and then discovering that the closest parking spot is
available.  For the optimistic strategy this circumstance is impossible by
construction, while for the prudent strategy it happens with probability
close to $1-\rho_1$, i.e., approximately in 89\% of all realizations.
Another efficacy measure is the fraction of times a driver has to backtrack
to the end of the parked cars because there are no open spots available.
Fortunately for the driver, as $\lambda$ increases and the number of parked
cars similarly increases, it becomes less likely that a parking attempt
requires backtracking to the end of the parking lot.

With regard to parking, humans do not follow optimal
strategies~\cite{traffic08}.  Instead, drivers tend to use simple heuristics,
and the strategies outlined in this work are examples of such heuristics.
Devising an optimal strategy is still an intriguing challenge.  Adapting the
methods from the optimal stopping research is not straightforward, e.g., in
the parking problem studied in \cite{Miller} the probability of a parking
spot being occupied is independent of its location or of whether neighboring
places are occupied.  In our problem, the spatial distribution is emergent,
and it also only statistically stationary.

\section*{Acknowledgments}
  PLK thanks the hospitality of the Santa Fe Institute where this work was
  completed.  SR gratefully acknowledges financial support from NSF grant
  DMR-1608211.  We also thank John Miller for helpful advice and
  conversations on this problem.

\bigskip
\bigskip
\newcommand{\newblock}{}

\end{document}